\documentclass{article}

\usepackage{arxiv}

\usepackage{hyperref}       
\usepackage{url}            
\usepackage{booktabs}       
\usepackage{amsfonts}       
\usepackage{nicefrac}       
\usepackage{microtype}      
\usepackage{lipsum}		
\usepackage{graphicx}
\usepackage{natbib}
\usepackage{doi}
\usepackage{amsmath}
\usepackage{fontspec}
\usepackage{polyglossia}
\usepackage{enumitem}
\usepackage{xcolor}

\setdefaultlanguage{english}

\newfontfamily\devanagarifont{NotoSansDevanagari}[
    Path=./,
    Extension=.ttf,
    UprightFont=*-Regular
]

\title{When De-noising Hurts: A Systematic Study of Speech Enhancement Effects on Modern Medical ASR Systems}

\author{
    Sujal Chondhekar \\
    \And
    Vasanth Murukuri \\
    \And    
    Rushabh Vasani \\
    \And   
    Sanika Goyal \\
    \And       
    Rajshree Badami \\
    \And    
    Anushree Rana \\
    \And    
    Sanjana SN \\
    \And
    Karthik Pandia \\
    \And
    Sulabh Katiyar \\
    \And    
    Neha Jagadeesh \\
    \And        
    Sankalp Gulati \\
    \And
    \centerline{\textbf{EkaCare (Orbi Health Private Limited), Bengaluru, India (sankalp@eka.care)}}   
}

\renewcommand{\headeright}{Technical Report}
\renewcommand{\undertitle}{Technical Report}

\date{}

\begin{document}

\maketitle
\begin{abstract}
Speech enhancement methods are commonly believed to improve the performance of automatic speech recognition (ASR) in noisy environments. However, the effectiveness of these techniques cannot be taken for granted in the case of modern large-scale ASR models trained on diverse, noisy data. We present a systematic evaluation of MetricGAN-plus-voicebank denoising on four state-of-the-art ASR systems—OpenAI Whisper, NVIDIA Parakeet, Google Gemini Flash 2.0, Parrotlet-a using 500 medical speech recordings under nine noise conditions. ASR performance is measured using semantic WER (semWER), a normalized word error rate (WER) metric accounting for domain-specific normalizations. Our results reveal a counterintuitive finding: speech enhancement preprocessing degrades ASR performance across all noise conditions and models. Original noisy audio achieves lower semWER than enhanced audio in all 40 tested configurations (4 models $\times$ 10 conditions), with degradations ranging from 1.1\% to 46.6\% absolute semWER increase. These findings suggest that modern ASR models possess sufficient internal noise robustness and that traditional speech enhancement may remove acoustic features critical for ASR. For practitioners deploying medical scribe systems in noisy clinical environments, our results indicate that preprocessing audio with noise reduction techniques might not just be computationally wasteful but also be potentially harmful to the transcription accuracy.
\end{abstract}


\section{Introduction}

The deployment of automatic speech recognition (ASR) for clinical documentation in India presents significant challenges due to high ambient noise levels. Hospitals often experience high noise levels from medical equipment, conversations, and infrastructure \citep{HospitalNoise2020}, far exceeding the WHO-recommended 40 dB limit for healthcare facilities \footnote{https://cpcb.nic.in/who-guidelines-for-noise-quality/}. Conventional wisdom suggests that speech enhancement preprocessing should improve ASR accuracy by removing noise and improving signal quality \citep{1318467}.

However, this assumption originates from an era when ASR systems used Gaussian Mixture Models (GMMs) and Hidden Markov Models (HMMs) trained on clean speech \citep{classical_asr}. Modern end-to-end neural ASR models like Whisper \citep{radford2022whisper}, trained on hundreds of thousands of hours of diverse, real-world audio including noisy recordings, may possess fundamentally different noise robustness characteristics. If these models have learned robust internal representations that handle noise, external preprocessing could potentially remove useful acoustic information.


This study is motivated by practical observations from real-world deployments of EkaScribe\footnote{https://ekascribe.ai/}, with the majority being done on non-mobile devices having typically a single microphone. The audio is captured through an operating system (OS) managed audio pipeline, which often applies signal processing such as noise suppression or echo cancellation by default. These speech enhancement mechanisms are designed to improve human listening quality, but whether they always improve ASR performance is not guaranteed. This raises an important practical concern: if modern ASR models are already robust to noise, such preprocessing may be unnecessary or even counterproductive. Understanding how explicit denoising interacts with modern ASR therefore becomes critical for deployment scenarios.

We address this gap through a systematic evaluation of MetricGAN-plus-voicebank denoising \citep{fu2021metricganplus} a state-of-the-art speech enhancement model on four leading ASR systems: OpenAI Whisper, NVIDIA Parakeet, Google Gemini Flash 2.0, and Parrotlet-a, a model specifically trained for English speech recognition in Indian healthcare. We tested 500 medical recordings under nine noise conditions representing typical Indian clinical environments. We release our evaluation code, dataset and detailed results to enable reproduction and extension of this work. These resources are publicly available on GitHub\footnote{https://github.com/eka-care/when-denoising-hurts}.



\section{Related Work}

\subsection{Modern ASR Systems}

Recent ASR advances stem from self or weakly-supervised learning on massive datasets. Whisper \citep{radford2022whisper} was trained on 680,000 hours of multilingual web audio, explicitly including diverse recording conditions and background noise. This exposure to real-world acoustic variability may confer inherent noise robustness. Similarly, NVIDIA's Parakeet models \citep{rekesh2023fastconformer,xu2023tdt} leverage large-scale training with data augmentation including noise injection. Google's Gemini 2.0 \citep{geminiteam2024} processes audio through multimodal encoders pre-trained on millions of examples. 

A critical distinction exists between these models and classical ASR systems: modern models learn end-to-end mappings from raw audio to text without explicit acoustic modelling assumptions. This may enable them to extract robust features from noisy signals without external preprocessing.

\subsection{Speech Enhancement Techniques}

Traditional speech enhancement techniques such as spectral subtraction and Wiener filtering were developed to improve human speech intelligibility and classical ASR mainly trained on clean data \citep{Loizou2007,Benesty2006}. These methods typically operate in the frequency domain, estimating and removing noise to recover clean speech signals.

Deep learning approaches including Deep Neural Networks (DNNs) and Generative Adversarial Networks (GANs) have largely superseded classical methods. MetricGAN \citep{fu2019metricgan} aims to directly optimize perceptual quality metrics (PESQ, STOI) rather than reconstruction error. The improved MetricGAN+ variant \citep{fu2021metricganplus} achieves state-of-the-art performance on standard benchmarks.

However, a critical gap exists: most speech enhancement evaluations focus on \textit{speech quality} metrics or performance of \textit{classical} ASR systems. Systematic evaluation of modern large-scale neural ASR remains a topic of research.

\subsection{Speech Enhancement Impact on ASR}

The relationship between speech enhancement and ASR accuracy has been repeatedly re-examined as ASR architectures evolved, yet consensus remains elusive.

\textbf{Classical ASR Era:} Traditional GMM-HMM systems trained on clean speech benefited substantially from preprocessing. \citet{797809} demonstrated that combined speech enhancement and auditory modeling improved distributed speech recognition in noisy conditions. These findings established the paradigm that enhancement should precede recognition, an assumption that persists in many deployment pipelines today.

\textbf{Deep Learning Transition:} The advent of deep neural networks for acoustic modelling prompted fundamental questions about preprocessing necessity. \citet{Delcroix2013} posed a prescient question in 2013: "Is speech enhancement pre-processing still relevant when using deep neural networks for acoustic modelling?" Their investigation of DNN-based acoustic models suggested that enhancement benefits were diminishing as models became more sophisticated. \citet{8462581} explored GAN-based enhancement for robust speech recognition, finding mixed results that varied by noise type and training conditions.

\textbf{Modern ASR Era:} Recent work reveals increasingly complex patterns. \citet{9414027} investigated end-to-end models for robust speech recognition, finding that modern architectures exhibit surprising noise robustness even without enhancement. \citet{9746489} analyzed noise suppression losses and discovered a critical trade-off: aggressive noise reduction can introduce speech distortion that harms ASR performance more than the original noise. This suggests that enhancement optimized for perceptual quality may not align with ASR requirements.

Recent comparative studies have revealed important nuances in ASR performance and the enhancement effect. \cite{agarwal2025robustness} demonstrated that Whisper exhibits significant sensitivity to background noise and domain shifts, while Gemini's multimodal architecture shows enhanced robustness through contextual adaptation. 
\citet{trabelsi2024noise} directly evaluated whether noise reduction improves open-source ASR engines, finding that enhancement effectiveness varies significantly across models and conditions. \citet{10884312} introduced DENOASR, revealing that denoising can inadvertently introduce or amplify demographic biases in ASR systems through selective acoustic filtering. Most recently, \citet{Nasretdinov_2025} proposed Schrödinger Bridge-based enhancement specifically designed for ASR, suggesting that traditional enhancement methods may be fundamentally misaligned with neural recognition objectives.

\textbf{The Current Gap:} While recent studies have begun to examine noise robustness in modern ASR systems, the effectiveness of commonly used speech enhancement pipelines in this context remains insufficiently characterized. Prior work has often focused on limited model scales, specific noise types, or perceptual quality metrics, and has produced mixed or model-dependent findings. In particular, there is a lack of systematic evaluation of how widely adopted enhancement methods interact with large-scale ASR models—such as Whisper—that are trained on extensive, noisy data, especially in domain-specific settings like medical speech. Given the practical importance of deployment in noisy and resource-constrained clinical environments, a careful empirical study under controlled noise conditions is needed.

In this work, we address this gap by evaluating a widely used enhancement method (MetricGAN+) on four modern ASR systems using medical speech and noise conditions representative of real-world clinical environments.

\section{Methodology}

\subsection{Research Question}

We investigate: \textit{Does MetricGAN-plus-voicebank speech enhancement improve ASR accuracy for modern models on noisy medical recordings?}

Our null hypothesis is that applying speech enhancement techniques like MetricGAN-plus-voicebank does not produce a statistically significant change in ASR accuracy compared to transcribing the original noisy audio.

\subsection{Dataset and Noise Conditions}

We compiled 500 annotated medical audio recordings from mocked clinical consultation settings, professionally transcribed to establish ground truth. These were mock consultations by the employees of EkaCare. The chosen samples were primarily in English to avoid adding another variable that could potential degrade the ASR performance. To simulate realistic hospital acoustics, we synthetically added three noise types to the original audio files at varying intensities using \href{https://github.com/iver56/audiomentations}{Audiomentations} library:

\textbf{Background Noise:} Continuous ambient sound (HVAC, equipment hum, distant conversations, these noises were taken from \href{https://www.kaggle.com/datasets/nafin59/hospital-ambient-noise}{hospital-ambient-noise-dataset}) at Signal-to-Noise Ratios (SNR) of 10, 30, and 50 dB.

\textbf{Short Noise:} Transient bursts (equipment beeps, door slams, nearby conversations, these were taken from \href{https://pixabay.com/sound-effects/search/short/}{pixabay}) at SNR 10, 30, and 50 dB.

\textbf{Gaussian Noise:} Additive white Gaussian noise at amplitudes 0.001, 0.009, and 0.017.

SNR is defined as:
\begin{equation}
\text{SNR}_{\text{dB}} = 10 \log_{10}\left(\frac{P_{\text{signal}}}{P_{\text{noise}}}\right)    
\end{equation}
where $P_{\text{signal}}$ and $P_{\text{noise}}$ represent speech and noise power over non-silent regions.

SNR 10dB corresponds to severe noise typical of busy clinical areas, 30 dB to moderate noise, and 50 dB to mild background noise. This yielded 4,500 noisy test utterances (500 recordings $\times$ 9 conditions). This dataset comprising original recordings, noise-augmentations and their de-noised versions, is publicly available on HuggingFace\footnote{https://huggingface.co/datasets/ekacare/denoising-impact-evaluation-dataset}.

\subsection{ASR Systems}

\textbf{OpenAI Whisper Large-v3} \citep{radford2022whisper}: Encoder-decoder Transformer trained on 680,000 hours of multilingual data. Processes 80-channel log-Mel spectrograms. We used the Large-v3 variant (1.55B parameters) that is publicly available on HuggingFace\footnote{https://huggingface.co/openai/whisper-large-v3}, hosted locally.

\textbf{NVIDIA Parakeet-TDT-1.1B} \citep{rekesh2023fastconformer,xu2023tdt}: Fast Conformer encoder with Token-and-Duration Transducer decoder trained on 64,000+ hours. Uses 8x depthwise-separable convolution downsampling. We use the model version publicly available on HuggingFace\footnote{https://huggingface.co/nvidia/parakeet-tdt-1.1b}, hosted locally via NeMo.

\textbf{Google Gemini Flash 2.0} \citep{geminiteam2024}: Multimodal model with native audio understanding, accessed via Gemini API.

\textbf{Parrotlet-a}: This is a purpose-built automatic speech recognition (ASR) model specifically trained for English speech in Indian healthcare settings, optimized for transcribing medical speech. This model combines the Whisper V3 large encoder and MedGemma-3-4B decoder through a lean projector layer for efficient speech-to-text conversion. Model description and release details are available here\footnote{https://info.eka.care/services/parrotlet-a-en-5b-releasing-our-purpose-built-llm-for-english-asr-in-indian-healthcare}.

\subsection{Denoising Method}

For denoising we used SpeechBrain's \citep{ravanelli2021speechbrain} MetricGAN-plus-voicebank model \citep{fu2021metricganplus}. This GAN-based approach optimizes perceptual quality (PESQ, STOI) rather than reconstruction error:

\begin{enumerate}
    \item Compute Short-Time Fourier Transform: $X[n,k] = \text{STFT}\{x[n]\}$
    \item Generate spectral mask via trained generator: $M[n,k] = G_{\theta}(|X[n,k]|)$
    \item Apply mask to noisy magnitude: $\hat{S}[n,k] = M[n,k] \cdot |X[n,k]|$
    \item Reconstruct via inverse STFT with original phase: $\hat{s}[n] = \text{ISTFT}\{\hat{S}[n,k] \cdot e^{j\angle X[n,k]}\}$
\end{enumerate}

All audio was resampled to 16 kHz mono before processing.

\subsection{Evaluation Methodology}

We employ a comprehensive evaluation process to assess ASR performance under noisy and enhanced conditions, structured as follows:
\begin{enumerate}
    \item \textbf{Baseline Establishment}
    \begin{itemize}
        \item Transcribed 500 original audio files (without added noise)
        \item Applied MetricGAN+ enhancement to 500 original files and transcribed them
    \end{itemize}
    \item \textbf{Noise Corruption and Dual Transcription}
    \begin{itemize}
        \item Corrupted each of the 500 original files with 9 noise conditions (3 types $\times$ 3 configurations)
        \item Generated 4,500 noisy audio files
        \item Each noisy file was transcribed under two conditions:
        \begin{itemize}
            \item \textbf{Noisy Condition}: Direct transcription without enhancement
            \item \textbf{Enhanced Condition}: Transcription after MetricGAN+ enhancement
        \end{itemize}
    \end{itemize}
    \item \textbf{Transcription Volume per ASR System}
    \begin{itemize}
     \item Baseline transcriptions: $500 + 500 = 1,000$
     \item Noisy condition transcriptions: $4,500$
    \item Enhanced condition transcriptions: $4,500$
    \item \textbf{Total transcriptions per ASR system}: $1,000 + 4,500 + 4,500 = 10,000$
    \end{itemize}
\end{enumerate}
\textbf{Total across all 4 ASR systems}: $4 \times 10,000 = 40,000$ transcriptions

This extensive dataset enables a thorough analysis of noise and enhancement impacts on each ASR system.

Word Error Rate (WER) was computed as:
\begin{equation}
\text{WER} = \frac{S + D + I}{N} \times 100\%
\end{equation}
where $S$ is substitutions, $D$ is deletions, $I$ is insertions, and $N$ is total reference words. Text normalization (lowercase, punctuation removal) was applied uniformly.

\textbf{Semantic WER (semWER)} \citep{karma-medeval2025}: Indian medical consultations frequently exhibit code-mixing—alternating between English medical terminology and regional languages (Hindi, Tamil, Telugu, etc.) mid-conversation. Standard WER penalizes semantically equivalent expressions. We computed semWER by applying additional normalizations:

\begin{enumerate}[leftmargin=*]
    
    \item \textbf{Unit Normalisations}
    \begin{itemize}
        \item Expand abbreviated units to their word forms or full representations
        \item Example: \texttt{5mg} $\rightarrow$ \texttt{five mg}, {\devanagarifont फाइव मिलीग्राम}, \texttt{five milligrams}
        \item Example: \texttt{2°C} $\rightarrow$ \texttt{2 degree C}, {\devanagarifont 2 डिग्री सेल्सियस}
    \end{itemize}
    
    \item \textbf{Symbol Normalisations}
    \begin{itemize}
        \item Convert mathematical and special symbols to their word equivalents
        \item Example: \texttt{3/4} $\rightarrow$ \texttt{3 by 4}, \texttt{3 per 4}, \texttt{three per four}
    \end{itemize}
    
    \item \textbf{Contraction Expansion}
    \begin{itemize}
        \item Expand contracted forms to their full word forms
        \item Example: \texttt{I'm} $\rightarrow$ \texttt{I am}, \texttt{It's} $\rightarrow$ \texttt{it is}
    \end{itemize}
    
    \item \textbf{Number to Word Conversion} \citep{indic-numtowords2025}
    \begin{itemize}
        \item Convert numeric digits to their word representations in various languages and scripts 
        \item Example: \texttt{150} $\rightarrow$ \texttt{one hundred and fifty}, {\devanagarifont एक सौ पचास}, {\devanagarifont १५०}
    \end{itemize}
    
    \item \textbf{Punctuation Removal} \citep{kunchukuttan2020indicnlp}
    \begin{itemize}
        \item Remove all punctuation marks from text including commas, periods, etc.
    \end{itemize}
    
    \item \textbf{Unicode Normalization}
    \begin{itemize}
        \item Convert all text to canonical Unicode form (NFC/NFKC) and merge equivalent codepoints
        \item Example: Merge composed vs. decomposed forms of matras in Indic scripts
    \end{itemize}
    
    \item \textbf{Character Canonicalization} \citep{kunchukuttan2020indicnlp}
    \begin{itemize}
        \item Standardize multiple codepoints that visually represent the same character to a single canonical form
        \item Example: Fix canonical order of combining marks ({\devanagarifont ं + ा})
    \end{itemize}
    
    \item \textbf{Nukta Variant Removal} \citep{kunchukuttan2020indicnlp}
    \begin{itemize}
        \item Replace nukta (dot below) forms with their base character equivalents
        \item Example: {\devanagarifont क़} $\rightarrow$ {\devanagarifont क}, {\devanagarifont फ़} $\rightarrow$ {\devanagarifont फ}
    \end{itemize}
    
    \item \textbf{Diacritic Normalization} \citep{kunchukuttan2020indicnlp}
    \begin{itemize}
        \item Standardize combining diacritical marks to ensure consistent order and convert variant forms to canonical representations
    \item Example: Normalize anusvara ({\devanagarifont कं}), chandrabindu ({\devanagarifont कँ}), visarga ({\devanagarifont कः})        
    \end{itemize}
    \item \textbf{Transliteration} \citep{googletransliteration2025}
    \begin{itemize}
        \item Convert text from one script to another while preserving phonetic pronunciation
        \item Example: \texttt{computer} $\rightarrow$ {\devanagarifont कंप्यूटर}
    \end{itemize}
\end{enumerate}

While semWER is designed to handle multilingual and code-mixed scenarios common in Indian medical consultations, since the dataset considered is English, only a subset of the normalisations mentioned above (unit expansion, number-to-word conversion, contractions, and symbol normalization in English) would have been relevant.

This produces a more clinically meaningful error metric for Indian healthcare ASR. All reported results use semWER unless otherwise noted.

\section{Results}

\subsection{ASR Accuracy and Impact of Noise}
\begin{figure}[htbp]
    \centering
    \fbox{\includegraphics[width=0.95\linewidth]{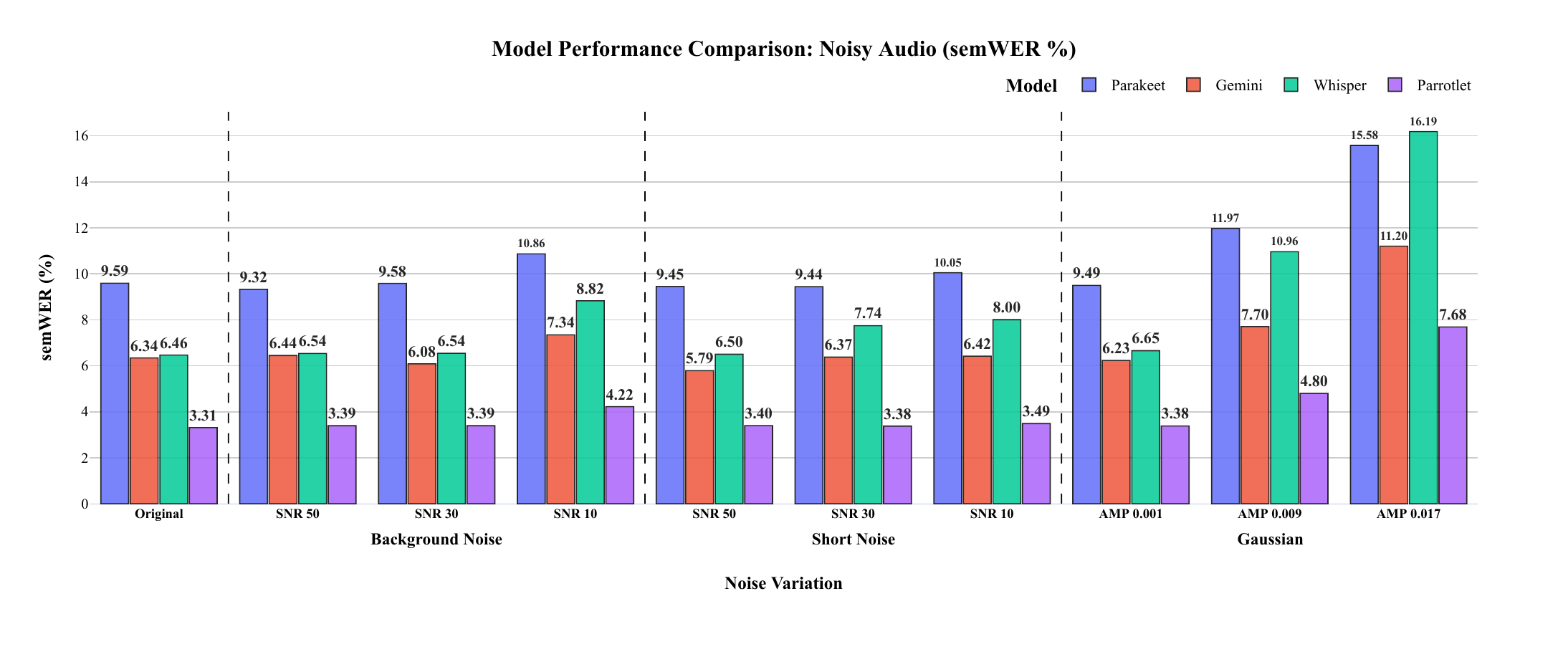}}
    \caption{semWER (\%) (lower is better) performance of all evaluated ASR models under various noisy conditions. Parrotlet-a achieves the lowest error rates across all conditions.}
    \label{fig:noisy_performance_all}
\end{figure}

\begin{figure}[htbp]
\centering
\fbox{\includegraphics[height=0.8\textheight]{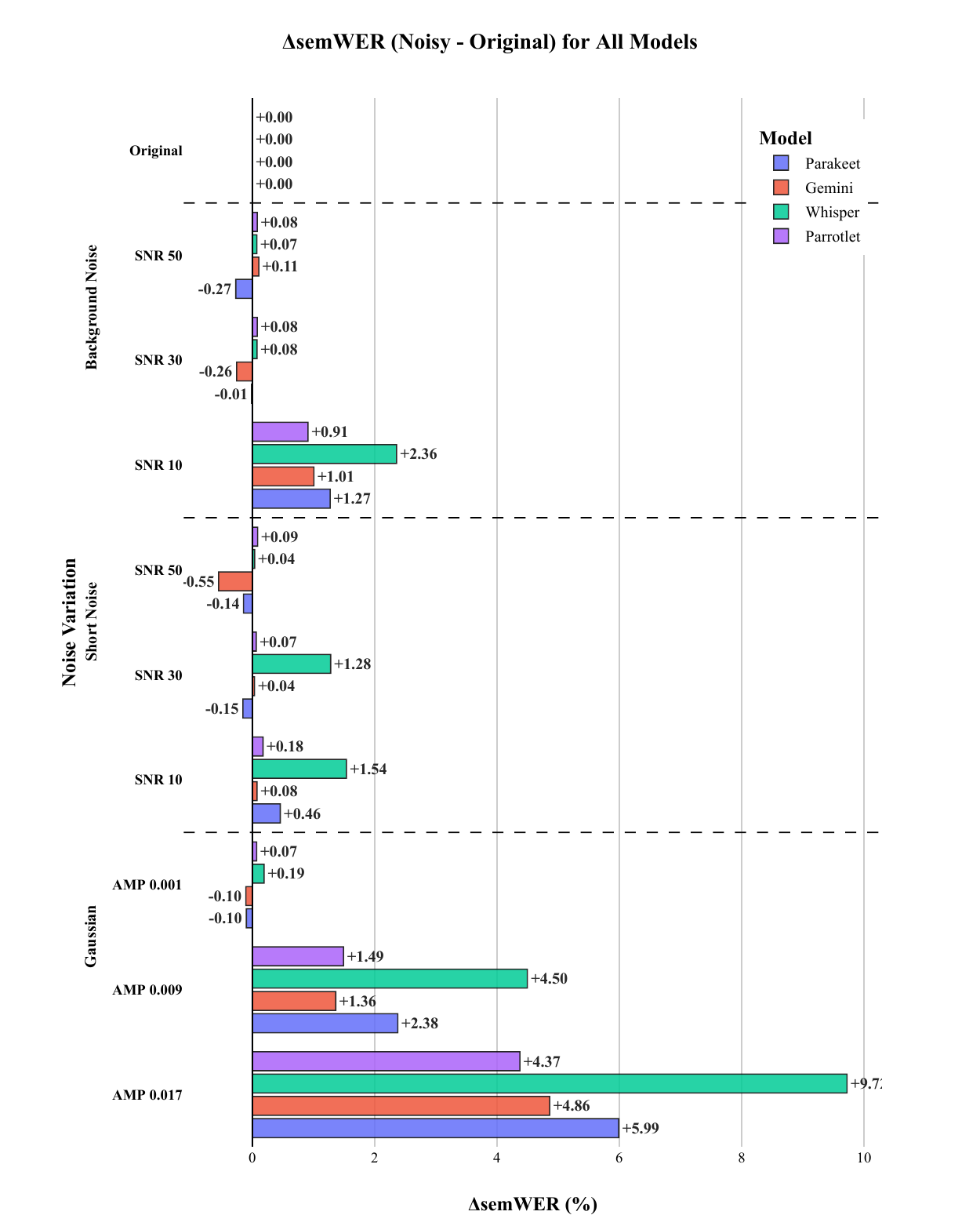}}
\caption{Performance of four ASR models in various conditions. We observe degradation in the performance of these models beyond certain SNR levels.}
\label{fig:noise_impact}
\end{figure}

Figure \ref{fig:noisy_performance_all} presents the ASR performance of four state-of-the-art models on original recordings and their noisy variants. The figure shows the baseline semWER (\%) for each model, revealing that Parrotlet-a consistently achieves the lowest error rates across all conditions.
To easily analyze the impact of noise on performance, in Figure \ref{fig:noise_impact} we show the performance difference ($\Delta$semWER) of these models under various noise conditions compared with the original conditions. A notable initial observation is that high SNR noise addition (i.e., low noise levels) even leads to marginal improvements in ASR performance for some models. However, as noise levels increase, ASR performance degrades significantly, with the extent of degradation varying across models. Whisper exhibits the greatest performance decline, particularly under Gaussian noise conditions (reaching approximately 9.72\% $\Delta$semWER at AMP 0.017), followed by Parakeet with moderate degradation. In contrast, Gemini and Parrotlet-a demonstrates the most robust behaviour across all noise conditions, maintaining relatively stable performance even in highly noisy environments. This suggests that the architectures of Gemini and Parrotlet-a are better equipped to handle acoustic variability and noise-corrupted inputs compared to the other evaluated models.

\subsection{Main Finding: Denoising Degrades Performance}

\begin{figure}[htbp]
    \centering
    \fbox{\includegraphics[width=0.95\linewidth]{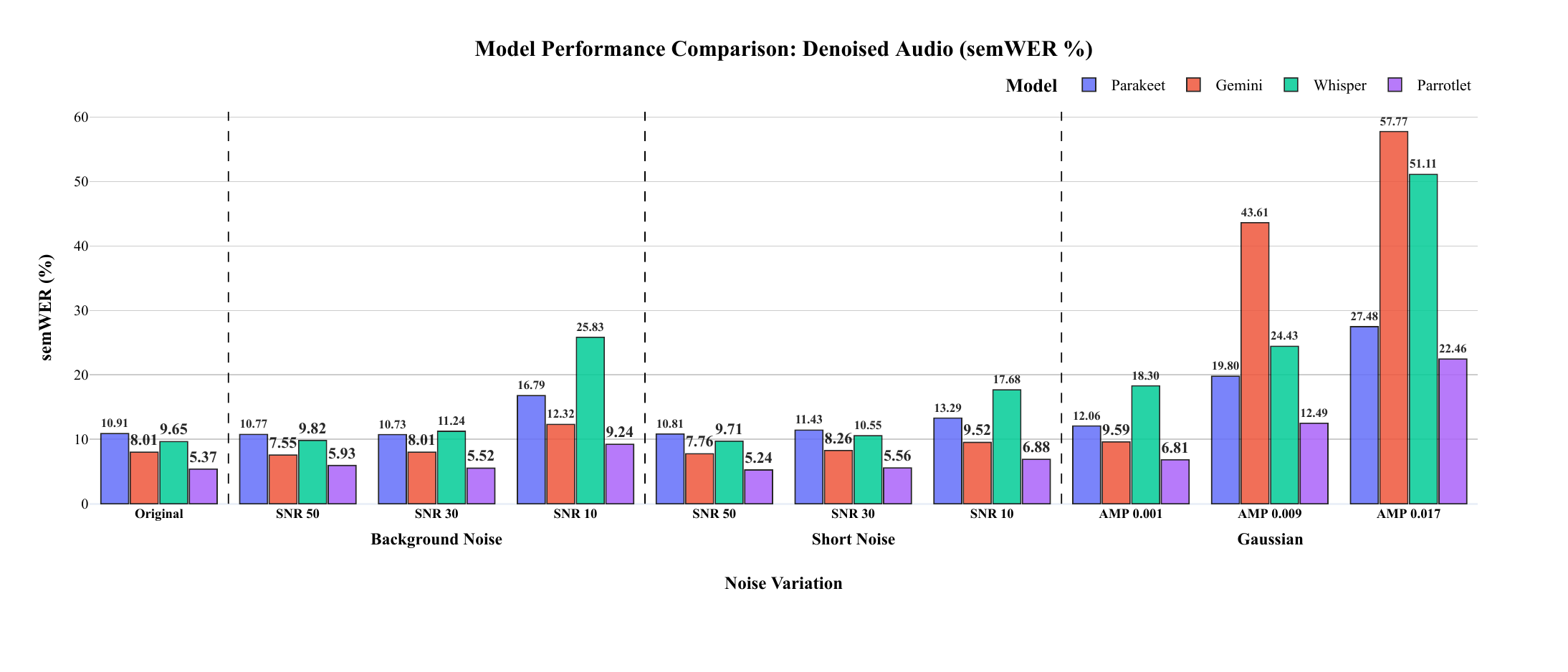}}
    \caption{semWER (\%) performance of all ASR models after enhancement with the SpeechBrain denoiser.}
    \label{fig:denoised_performance_all}
\end{figure}

The ASR performance on de-noised recordings using the SpeechBrain denoiser is visualized in Figure \ref{fig:denoised_performance_all}, which presents the semWER (\%) for all models after enhancement processing. Notably, Gemini and Whisper experience catastrophic degradation post denoising of certain noise conditions, with semWER soaring to extreme values (e.g., Gemini exceeding 50\% semWER), transforming them from competitively robust models to the worst performers. Parrotlet-a maintains its position as the most robust model overall, showing the smallest absolute error increases and the best final performance across all de-noised conditions

\begin{figure}[htbp]
\centering
\fbox{\includegraphics[height=0.8\textheight]{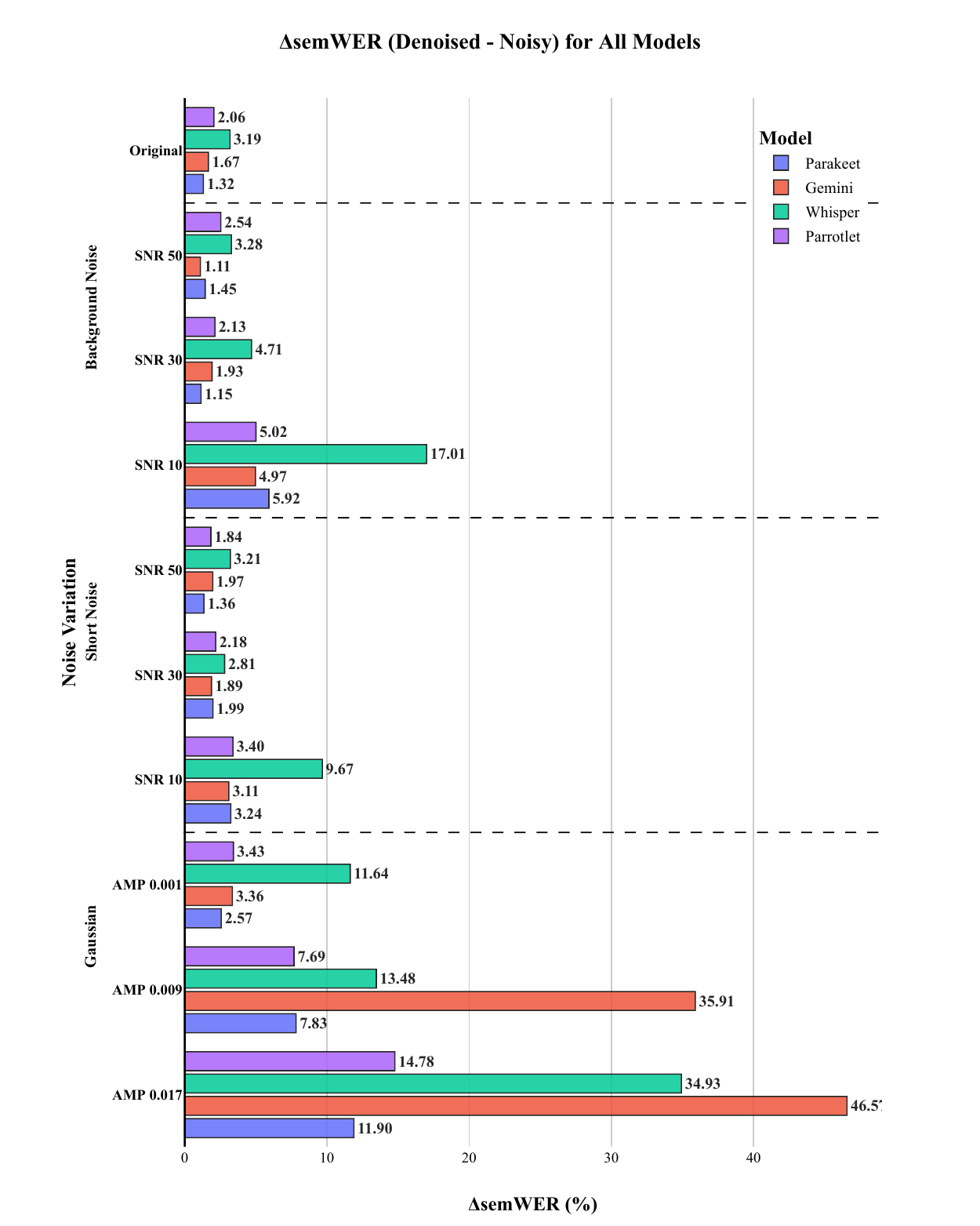}}
\caption{Change in Semantic Word Error Rate ($\Delta$semWER) after denoising across ASR models and noise conditions. We observe performance degradation due to denoising. Results demonstrate significant variation in denoising effectiveness across models, with Whisper showing the most substantial sensitivity to enhancement.}
\label{fig:denoising_impact}
\end{figure}

Figure \ref{fig:denoising_impact} shows the change in ASR performance resulting from the denoising process. It presents the $\Delta$semWER (De-noised $-$ Noisy) for all models, directly comparing the error rates for noisy audio against the same audio after enhancement. The positive $\Delta$semWER values indicate that de-noised audio yields higher semantic word error rates compared to processing the noisy audio directly.
The magnitude of performance degradation varies substantially across noise types and models. For moderate noise conditions (Background Noise SNR 10-50 and Short Noise SNR 10-50), the degradation remains relatively modest. However, the effect becomes dramatically more severe under Gaussian noise conditions. At higher Gaussian noise amplitudes (AMP 0.009 and 0.017), Gemini exhibits catastrophic performance collapse, with $\Delta$semWER values reaching 35.91\% and 46.57\% respectively. Whisper also shows substantial degradation at these noise levels (13.48\% and 34.93\%), Parakeet demonstrates more moderate but still significant decline (7.83\% and 11.9\%). Parrotlet-a also shows a similar pattern of degradation (7.69\% and 14.78\%).
Notably, even in original conditions, denoising introduces measurable performance penalties across all models (1.32\% to 3.19\%), suggesting that the denoising process itself introduces artifacts or information loss that adversely affects ASR accuracy. This finding challenges the conventional assumption that speech enhancement preprocessing universally benefits downstream ASR tasks, and indicates that modern ASR models may be better equipped to handle noisy inputs directly rather than relying on de-noised representations.

\subsection{Detailed Analysis by Noise Type}

\begin{figure}[htbp]
\centering
\fbox{\includegraphics[width=0.95\linewidth]{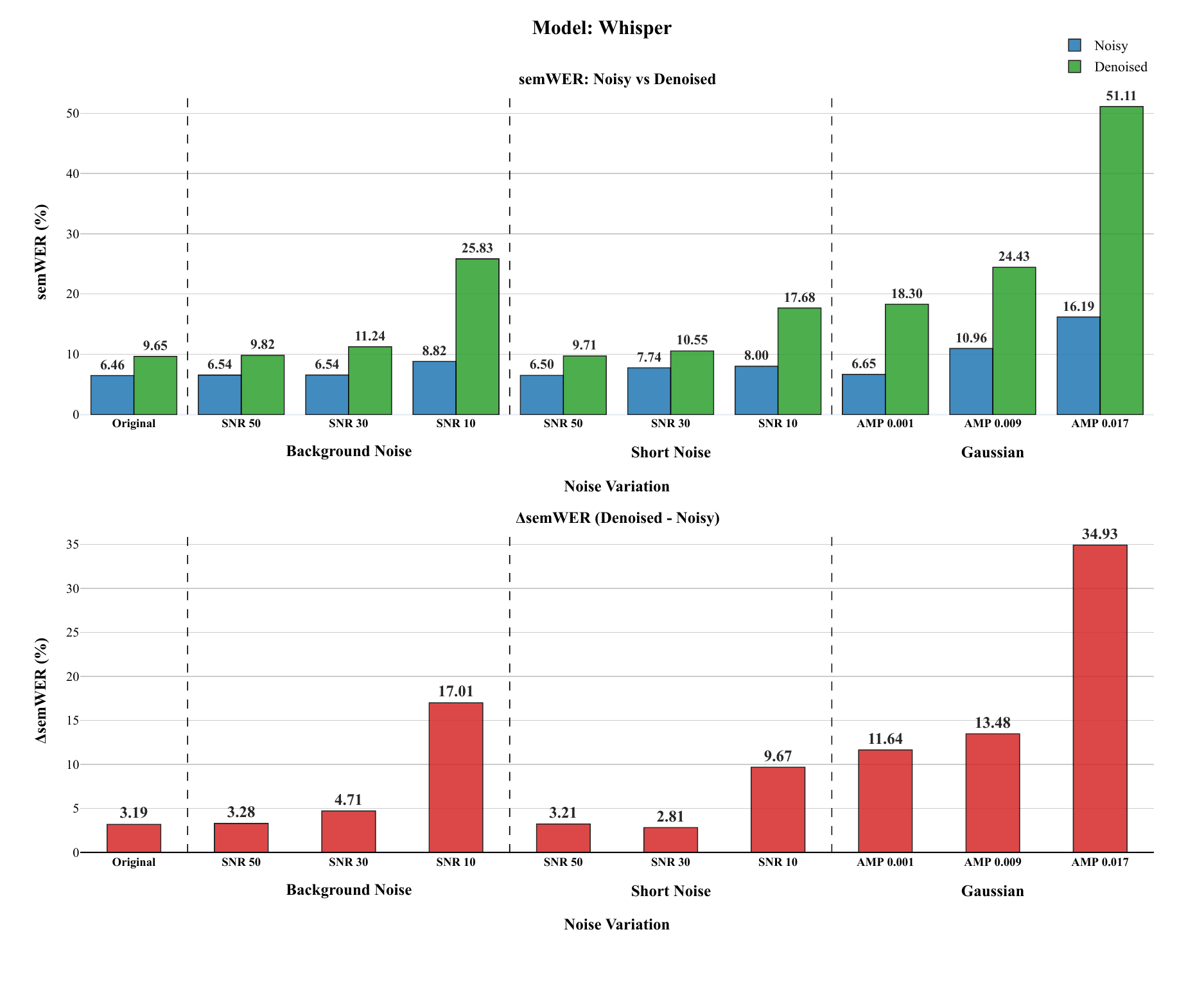}}
\caption{Whisper semWER (\%) and $\Delta$semWER. Enhancement consistently increases semWER across all conditions, with extreme degradation under Gaussian noise.}
\label{fig:whisper}
\end{figure}

\textbf{Background Noise:} Figure \ref{fig:whisper} shows Whisper's performance. At SNR 10dB (severe noise), noisy semWER is 8.82\% while enhanced semWER is 25.83\%—a catastrophic 17.0 percentage point degradation. Even at SNR 50 dB (mild noise), enhancement increases semWER from 6.54\% to 9.82\%.

\begin{figure}[htbp]
\centering
\fbox{\includegraphics[width=0.95\linewidth]{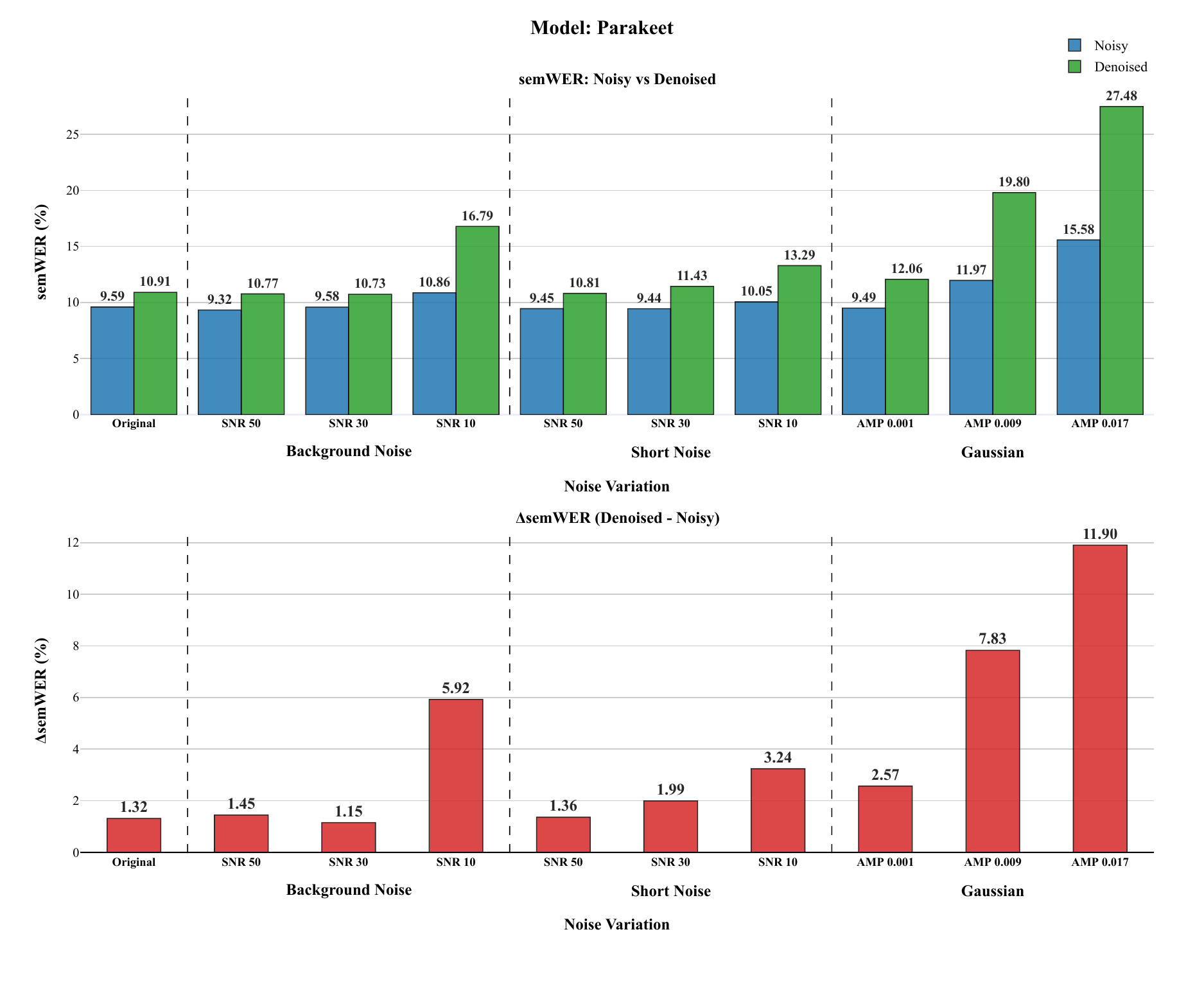}}
\caption{Parakeet semWER (\%) and $\Delta$semWER. Enhancement degrades performance across all noise conditions, though the magnitude is smaller than Whisper or Gemini. Parakeet maintains relatively consistent performance but still shows universal degradation from preprocessing.}
\label{fig:parakeet}
\end{figure}

\begin{figure}[htbp]
\centering
\fbox{\includegraphics[width=0.95\linewidth]{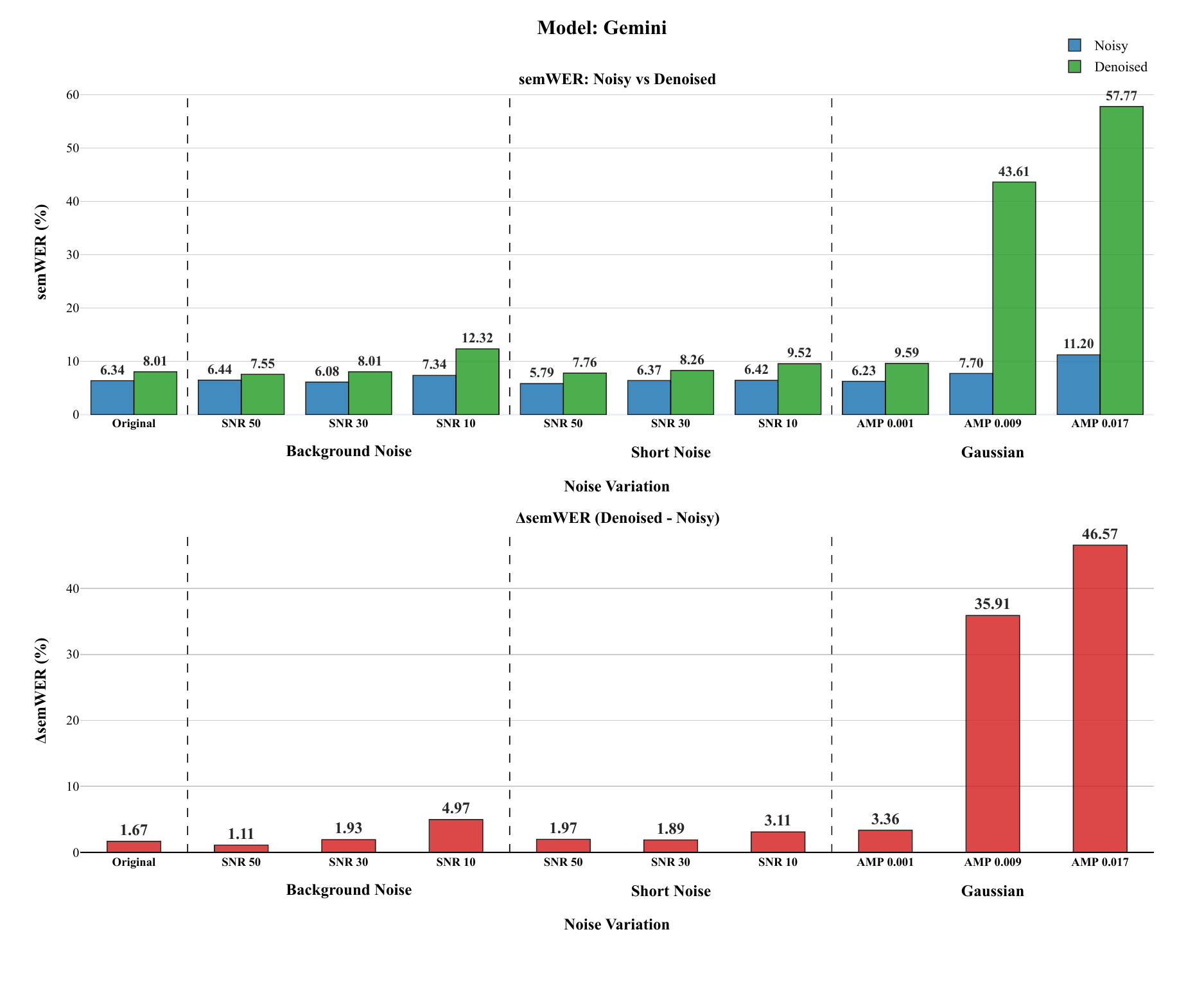}}
\caption{Gemini semWER (\%) and $\Delta$semWER. Enhancement dramatically degrades performance under Gaussian noise (semWER increases from 11\% to 58\% at amplitude 0.017). Other noise types show consistent but more modest degradation.}
\label{fig:gemini}
\end{figure}

\begin{figure}[htbp]
\centering
\fbox{\includegraphics[width=0.95\linewidth]{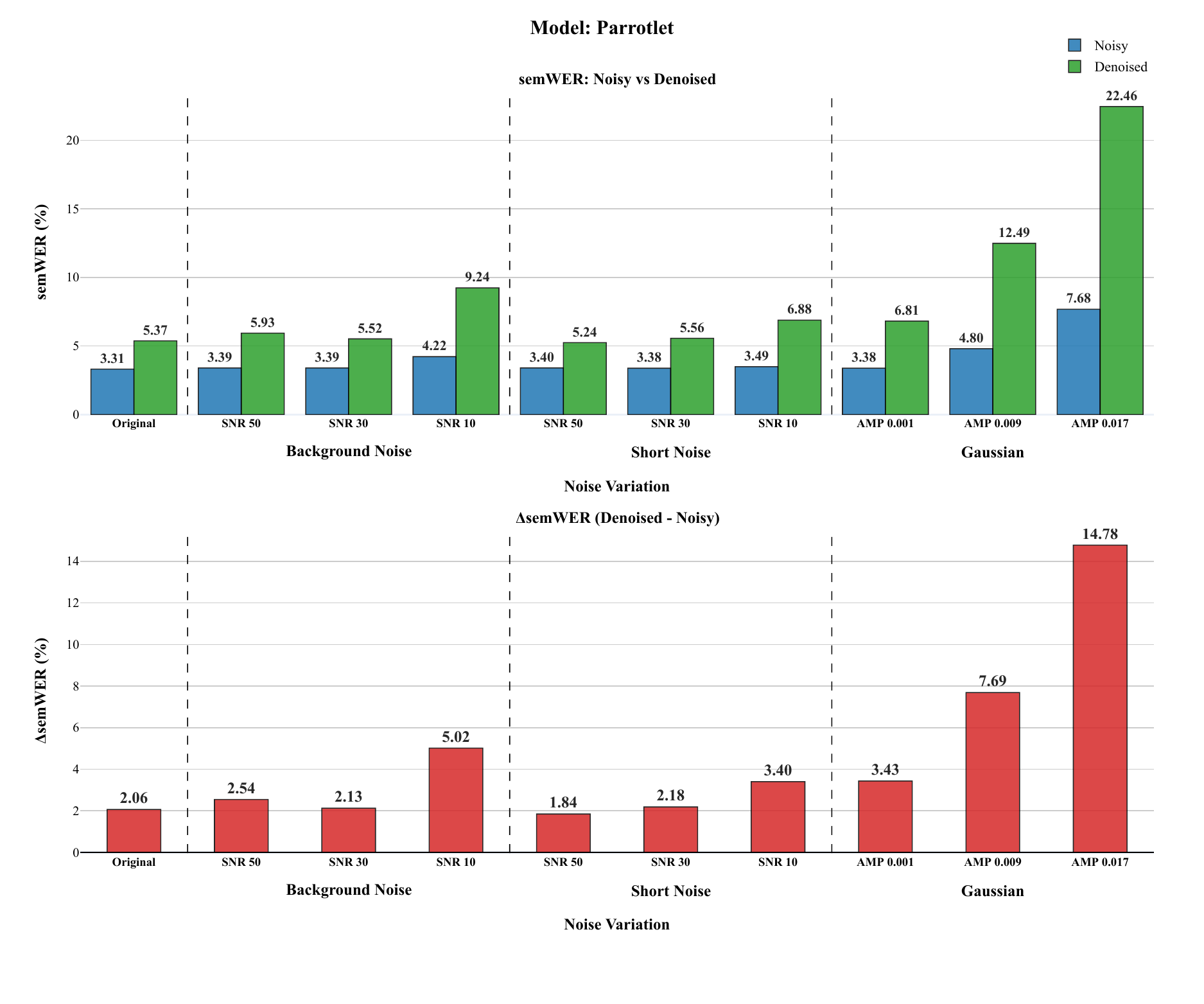}}
\caption{Parrotlet-a semWER (\%) and $\Delta$semWER. As compared to other model Parrotlet-a shows least semWER in noisy condition but denoising still degrade the performance.}
\label{fig:parrotlet}
\end{figure}

Parakeet shows similar patterns (Figure \ref{fig:parakeet}), though degradation is less severe. At SNR 10dB, semWER increases from 10.86\% to 16.79\%. Gemini (Figure \ref{fig:gemini}) shows moderate degradation: 7.34\% to 12.32\% at SNR 10dB, while Parrotlet-a (Figure \ref{fig:parrotlet}) exhibits relatively less degradation, increasing only from 4.22\% to 9.24\% at SNR 10dB.

\textbf{Short Noise:} Transient noise bursts show similar patterns. At SNR 10dB, Whisper's semWER increases from 8.0\% to 17.68\%. Parakeet: 10.05\% to 13.29\%. Gemini: 6.42\% to 9.52\%. Parrotlet-a: 3.49\% to 6.88\%. Enhancement provides no benefit at any SNR level for short noise.

\textbf{Gaussian Noise:} This condition produces the most dramatic degradation. At amplitude 0.017, Whisper's semWER increases from 16.19\% to 51.11\%, while Gemini shows an even more severe degradation from 11.2\% to 57.77\%. Parakeet also experiences a significant rise from 15.58\% to 27.48\%. Parrotlet-a also shows a similar degradation pattern from 7.69\% to 22.46\%. Even at low amplitude (0.001), enhancement still degrades performance for all models rather than helping.

\subsection{Quantitative Summary}
Table \ref{tab:delta} presents the  $\Delta$semWER statistics computed across all 40 test configurations. These configurations comprise 4 different ASR Models evaluated under 10 distinct noise conditions: one original (clean) condition and nine noisy conditions spanning three noise types at three intensity levels each. The $\Delta$semWER metric quantifies the change in semantic word error rate after applying speech enhancement: positive values indicate that enhancement worsened the model's performance.

\begin{table}[h]
\centering
\small 
\setlength{\tabcolsep}{4pt} 
\caption{$\Delta$semWER statistics. Positive values indicate enhancement increased semWER (degraded performance). ALL 40 configurations show degradation.}
\label{tab:delta}
\begin{tabular}{lcccccc}
\toprule
\textbf{Statistic} & \textbf{All Configs} & \textbf{Original} & \textbf{Background} & \textbf{Short} & \textbf{Gaussian} \\
\midrule
Configs with $\Delta$semWER > 0 & 40 / 40 & 4 / 4 & 12 / 12 & 12 / 12 & 12 / 12 \\
Mean $\Delta$semWER (\%) & +7.83 & +2.06 & +4.26 & +3.05 & +16.17 \\
Median $\Delta$semWER (\%) & +3.32 & +1.86 & +2.91 & +2.49 & +11.77 \\
Max degradation (\%) & +46.57 & +3.19 & +17.0 & +9.67 & +46.57 \\
Min degradation (\%) & +1.11 & +1.32 & +1.11 & +1.36 & +2.57 \\
\bottomrule
\end{tabular}
\end{table}

There are \textbf{zero configurations} where enhancement reduces semWER. The universality of this negative result suggests a fundamental incompatibility between MetricGAN+ enhancement and modern ASR models.

\section{Discussion}

Our results challenge the most common assumption that speech enhancement improves ASR. We present two main hypotheses:

\textbf{Hypothesis 1: Modern ASR Has Learned Internal Noise Robustness.} Whisper was explicitly trained on 680,000 hours of diverse, real-world audio including noisy recordings \citep{radford2022whisper}. The model may have learned robust internal representations that handle typical acoustic noise without external preprocessing. Enhancement may actually remove subtle acoustic cues (prosody, fine-grained spectral structure) that aid recognition.


\textbf{Hypothesis 2: Enhancement Artifacts.} Speech enhancement can introduce processing artifacts such as spectral smearing, temporal discontinuities and unnatural formant transitions. While these may be imperceptible to human listeners (and thus score well on PESQ/STOI), neural ASR models trained on natural speech may be sensitive to such artifacts.

\subsection{Generalizability and Limitations}

\textbf{Generalizability:} Our results may extend to other modern large-scale ASR models (e.g., wav2vec 2.0) trained on diverse data, though verification is needed. The finding may not generalize to:
\begin{itemize}
    \item Classical ASR systems (GMM-HMM) trained only on clean speech
    \item Small-scale models trained on limited data
    \item Extremely noisy conditions (SNR < 0 dB) beyond our test range
\end{itemize}

\textbf{Limitations:}  Our study has several constraints:

\textit{Single Enhancement Method:} We evaluated only MetricGAN-plus-voicebank model. Other approaches (Transformer-based enhancement, diffusion models) may behave differently.

\textit{Synthetic Noise:} We used synthetic methods of noise addition. While this enabled controlled comparisons, real clinical acoustics include reverberation and complex multi-source interference. Field validation is needed.

\textit{Dataset Size:} With 500 recordings, our sample size is smaller than ideal for establishing a generalisation on this topic. The combinatorial nature of the number of experiments kept us from increasing the number. 

\textit{Medical Domain:} Results may differ for other domains (conversational, broadcast). Medical speech with technical terminology may exhibit different noise robustness characteristics.

\textit{Indian English:} Our dataset comprises Indian English medical consultations. Generalization to other accents and languages requires verification.

\subsection{Future Directions}

\textbf{Alternative Enhancement Methods:} Evaluate modern approaches (e.g., Transformer-based enhancement, diffusion models) to determine if degradation is specific to GAN-based methods.

\textbf{End-to-End Joint Training:} Develop architectures that jointly optimize enhancement and recognition for ASR accuracy rather than perceptual quality.

\textbf{Analysis of Failure Modes:} Detailed investigation of which acoustic phenomena cause enhancement to fail—formant distortion, prosody removal, artefact introduction—could guide improvement.

\textbf{Real-World Deployment Study:} Pilot deployments in Indian hospitals with original noisy and enhanced audio to validate results in production environments.

\textbf{Multi-Microphone Approaches:} Evaluate beamforming and spatial filtering as alternatives to single-channel enhancement.

\section{Conclusion}

This study systematically evaluated the effect of a widely used speech enhancement method, MetricGAN-plus-voicebank, on four modern ASR systems in a noisy clinical environment. Across all tested noise conditions and models, we observed that this particular enhancement approach consistently increased word error rate relative to directly transcribing the noisy audio, consistently degrading performance.

Our results suggest that enhancement approaches designed to optimize human perceptual metrics may not align well with the representations learned by large-scale ASR models trained on noisy, real-world data.
Importantly, this does not imply that speech enhancement is inherently detrimental to ASR. Instead, it highlights that the effectiveness of enhancement is highly dependent on the specific technique employed, the noise characteristics, and the training methodology of the ASR model. Alternative approaches—such as ASR-aware enhancement, joint optimization, or domain-specific fine-tuning—may yield different outcomes and need further investigation.

From a practical viewpoint, our results suggest that denoising techniques like MetricGAN+ preprocessing should not be applied by default in medical ASR pipelines, and that its impact should be evaluated for each specific task and model. More broadly, these findings point to the need to re-examine common assumptions about audio preprocessing in the context of modern ASR systems that are already trained to operate under noisy conditions.


\bibliographystyle{plainnat}
\bibliography{references}

\begin{thebibliography}{25}
\providecommand{\natexlab}[1]{#1}
\providecommand{\url}[1]{\texttt{#1}}
\expandafter\ifx\csname urlstyle\endcsname\relax
  \providecommand{\doi}[1]{doi: #1}\else
  \providecommand{\doi}{doi: \begingroup \urlstyle{rm}\Url}\fi

\bibitem[Agarwal and Misra(2025)]{agarwal2025robustness}
Manav Agarwal and Anurag Misra.
\newblock Robustness and hallucination in asr: A comparative evaluation of whisper and gemini.
\newblock \emph{IOSR Journal of Computer Engineering}, 27\penalty0 (3):\penalty0 65--69, 2025.

\bibitem[{AI4Bh\=arat}(2025)]{indic-numtowords2025}
{AI4Bh\=arat}.
\newblock indic-numtowords: A python library to convert numbers to words for indian languages and english.
\newblock Python Package Index (PyPI), 2025.
\newblock URL \url{https://pypi.org/project/indic-numtowords/}.

\bibitem[Benesty et~al.(2006)Benesty, Makino, and Chen]{Benesty2006}
Jacob Benesty, Shoji Makino, and Jingdong Chen.
\newblock \emph{Speech Enhancement}.
\newblock Springer, 2006.

\bibitem[Braun and Gamper(2022)]{9746489}
Sebastian Braun and Hannes Gamper.
\newblock Effect of noise suppression losses on speech distortion and asr performance.
\newblock In \emph{IEEE ICASSP}, pages 996--1000, 2022.

\bibitem[Delcroix et~al.(2013)Delcroix, Kubo, Nakatani, and Nakamura]{Delcroix2013}
Marc Delcroix, Yusuke Kubo, Tomohiro Nakatani, and Atsushi Nakamura.
\newblock Is speech enhancement pre-processing still relevant when using deep neural networks for acoustic modeling?
\newblock In \emph{INTERSPEECH}, pages 2992--2996, 2013.

\bibitem[Donahue et~al.(2018)Donahue, Li, and Prabhavalkar]{8462581}
Chris Donahue, Bo~Li, and Rohit Prabhavalkar.
\newblock Exploring speech enhancement with generative adversarial networks for robust speech recognition.
\newblock In \emph{IEEE ICASSP}, pages 5024--5028, 2018.

\bibitem[Flynn and Jones(2008)]{797809}
Robert Flynn and Edward Jones.
\newblock Combined speech enhancement and auditory modelling for robust distributed speech recognition.
\newblock \emph{Speech Communication}, 50:\penalty0 797--809, 2008.

\bibitem[Fu et~al.(2019)Fu, Liao, Tsao, and Lin]{fu2019metricgan}
Szu-Wei Fu, Chien-Feng Liao, Yu~Tsao, and Shou-De Lin.
\newblock Metricgan: Generative adversarial networks based black-box metric scores optimization for speech enhancement.
\newblock In \emph{International Conference on Machine Learning (ICML)}, pages 2031--2041, 2019.

\bibitem[Fu et~al.(2021)Fu, Yu, Hsieh, Plantinga, Ravanelli, Lu, and Tsao]{fu2021metricganplus}
Szu-Wei Fu, Cheng Yu, Tsung-An Hsieh, Peter Plantinga, Mirco Ravanelli, Xugang Lu, and Yu~Tsao.
\newblock Metricgan+: An improved version of metricgan for speech enhancement.
\newblock \emph{arXiv preprint arXiv:2104.03538}, 2021.

\bibitem[{Gemini Team}(2024)]{geminiteam2024}
{Gemini Team}.
\newblock Gemini 1.5: Unlocking multimodal understanding across millions of tokens of context.
\newblock \emph{arXiv preprint arXiv:2403.05530}, 2024.

\bibitem[{Google}(2025)]{googletransliteration2025}
{Google}.
\newblock google-transliteration-api: A python interface for google's transliteration service.
\newblock Python Package Index (PyPI), 2025.
\newblock URL \url{https://pypi.org/project/google-transliteration-api/}.

\bibitem[Hinton et~al.(2012)]{classical_asr}
Geoffrey Hinton et~al.
\newblock Deep neural networks for acoustic modeling in speech recognition.
\newblock \emph{IEEE Signal Processing Magazine}, 29\penalty0 (6):\penalty0 82--97, 2012.

\bibitem[Joseph et~al.(2020)Joseph, Mehazabeen, and U]{HospitalNoise2020}
B.~E. Joseph, H.~Mehazabeen, and M.~U.
\newblock Noise pollution in hospitals - a study of public perception.
\newblock \emph{Noise \& Health}, 22\penalty0 (104):\penalty0 28--33, 2020.
\newblock \doi{10.4103/nah.NAH_13_20}.

\bibitem[Kunchukuttan(2020)]{kunchukuttan2020indicnlp}
Anoop Kunchukuttan.
\newblock The indicnlp library.
\newblock GitHub repository, 2020.
\newblock URL \url{https://github.com/anoopkunchukuttan/indic_nlp_library}.

\bibitem[Loizou(2007)]{Loizou2007}
Philipos~C. Loizou.
\newblock \emph{Speech Enhancement: Theory and Practice}.
\newblock CRC Press, 2007.

\bibitem[Nasretdinov et~al.(2025)Nasretdinov, Korostik, and Juki{\'c}]{Nasretdinov_2025}
Rustam Nasretdinov, Roman Korostik, and Ante Juki{\'c}.
\newblock Robust speech recognition with schr{\"o}dinger bridge-based speech enhancement.
\newblock In \emph{IEEE ICASSP}, pages 1--5, 2025.

\bibitem[Prasad et~al.(2021)Prasad, Jyothi, and Velmurugan]{9414027}
Anusha Prasad, Preethi Jyothi, and Raghuraman Velmurugan.
\newblock An investigation of end-to-end models for robust speech recognition.
\newblock In \emph{IEEE ICASSP}, pages 6893--6897, 2021.

\bibitem[Radford et~al.(2022)Radford, Kim, Xu, Brockman, McLeavey, and Sutskever]{radford2022whisper}
Alec Radford, Jong~Wook Kim, Tao Xu, Greg Brockman, Casey McLeavey, and Ilya Sutskever.
\newblock Robust speech recognition via large-scale weak supervision.
\newblock \emph{arXiv preprint arXiv:2212.04356}, 2022.

\bibitem[Rai et~al.(2024)Rai, Jaiswal, Prakash, Sree, and Mukherjee]{10884312}
A.~K. Rai, S.~D. Jaiswal, S.~Prakash, B.~P. Sree, and A.~Mukherjee.
\newblock Denoasr: Debiasing asrs through selective denoising.
\newblock In \emph{IEEE International Conference on Knowledge Graph (ICKG)}, pages 283--290, 2024.

\bibitem[Ravanelli et~al.(2021)]{ravanelli2021speechbrain}
Mirco Ravanelli et~al.
\newblock Speechbrain: A general-purpose speech toolkit.
\newblock \emph{arXiv preprint arXiv:2106.04624}, 2021.

\bibitem[Rekesh et~al.(2023)Rekesh, Koluguri, Kriman, Majumdar, Noroozi, Huang, Hrinchuk, Puvvada, Kumar, Balam, and Ginsburg]{rekesh2023fastconformer}
Dinesh Rekesh, Nithin~Rao Koluguri, Samuel Kriman, Somshubra Majumdar, Vahid Noroozi, Haicheng Huang, Oleksii Hrinchuk, Kishore Puvvada, Ankit Kumar, Jagadeesh Balam, and Boris Ginsburg.
\newblock Fast conformer with linearly scalable attention for efficient speech recognition.
\newblock In \emph{IEEE Automatic Speech Recognition and Understanding Workshop (ASRU)}, 2023.

\bibitem[Trabelsi et~al.(2024)Trabelsi, Werey, Warichet, and Helbert]{trabelsi2024noise}
Ahmed Trabelsi, Loic Werey, Sylvain Warichet, and Eric Helbert.
\newblock Is noise reduction improving open-source asr transcription engines quality?
\newblock In \emph{ICAART}, volume~3, pages 1221--1228, 2024.

\bibitem[Vasanth(2025)]{karma-medeval2025}
Vasanth.
\newblock karma-medeval: Evaluation framework for medical ai models.
\newblock Python Package Index (PyPI), 2025.
\newblock URL \url{https://pypi.org/project/karma-medeval/}.

\bibitem[Xu et~al.(2023)Xu, Jia, Majumdar, Huang, Watanabe, and Ginsburg]{xu2023tdt}
Hainan Xu, Feng Jia, Somshubra Majumdar, Haicheng Huang, Shinji Watanabe, and Boris Ginsburg.
\newblock Efficient sequence transduction by jointly predicting tokens and durations.
\newblock In \emph{International Conference on Machine Learning (ICML)}, pages 38491--38510, 2023.

\bibitem[Zhu and O'Shaughnessy(2003)]{1318467}
Wei Zhu and Douglas O'Shaughnessy.
\newblock Using noise reduction and spectral emphasis techniques to improve asr performance in noisy conditions.
\newblock In \emph{IEEE Workshop on Automatic Speech Recognition and Understanding (ASRU)}, pages 357--362, 2003.

\end{thebibliography}
\end{document}